\documentclass[12pt]{article}
\begin{document}

\def\mpd{\relax}
\def\mpu{\relax}
\def\mpl{\relax}

\newcommand{\bq}{\begin{equation}}
\newcommand{\eq}{\end{equation}}
\newcommand{\bqa}{\begin{eqnarray}}
\newcommand{\eqa}{\end{eqnarray}}
\newcommand{\nl}{\nonumber \\}
\newcommand{\mc}{Monte Carlo}
\newcommand{\qmc}{Quasi-Monte Carlo}
\newcommand{\qran}{quasi-random}
\newcommand{\intl}{\int\limits}
\newcommand{\intk}{\intl_K}
\newcommand{\suml}{\sum\limits}
\newcommand{\prol}{\prod\limits}
\newcommand{\umu}{^{\mu}}
\newcommand{\order}[1]{{\cal O}\left(#1\right)}
\newcommand{\eqn}[1]{Eq.(\ref{#1})}
\newcommand{\intinf}{\intl_{-\infty}^{\infty}}
\newcommand{\intinfi}{\intl_{-i\infty}^{i\infty}}
\newcommand{\ddf}{{\cal D}_{\!\!f}}
\newcommand{\si}{\sigma}
\newcommand{\vn}{\vec{n}}
\newcommand{\svn}{_{\vn}}
\newcommand{\sivn}{\si\svn}
\newcommand{\uvn}{u\svn}
\newcommand{\fal}[1]{^{\underline{#1}}}
\newcommand{\bpic}{\begin{picture}}
\newcommand{\epic}{\end{picture}}
\newcommand{\Cross}[4]{\Line(#1,#2)(#3,#4)\Line(#1,#4)(#3,#2)}
\newcommand{\DL}[2]{\DashLine(#1)(#2){1.5}}
\newcommand{\DC}[2]{\DashCArc(#1)(#2,0,360){2}}
\newcommand{\BC}{\BCirc}
\newcommand{\Yv}[1]{\BCirc(#1){3}}
\newcommand{\Vix}[1]{\Vertex(#1){2}}
\newcommand{\supm}{^{(m)}}
\newcommand{\om}{\omega}
\newcommand{\clt}{Central Limit Theorem}
\newcommand{\ezf}{\left\langle e^{zf(x)/N}\right\rangle}
\newcommand{\avg}[1]{\left\langle #1\right\rangle}


\pagestyle{empty}
\begin{flushright}
NIKHEF 96-020 \\
THEF-NYM-96.03
\end{flushright}

\begin{center}
\begin{Large}
{\bf Discrepancy-based error estimates\\
 for \qmc.\\
\vspace{\baselineskip}
III: Error distributions and central limits}\end{Large}\\
\vspace{\baselineskip}
{\bf Jiri Hoogland
  \footnote{e-mail: t96@nikhefh.nikhef.nl, 
   research supported by Stichting FOM}\\
  NIKHEF-H, Amsterdam, The Netherlands}\\
{\bf and \\}
{\bf Ronald Kleiss\footnote{e-mail: kleiss@sci.kun.nl}\\
University of Nijmegen, Nijmegen, The Netherlands}\\
\vspace{2\baselineskip}
{\bf Abstract}
\end{center}
In \qmc\ integration, the integration error is believed to be
generally smaller than in classical \mc\ with the same number of
integration points. Using an appropriate definition of an ensemble
of \qran\ point sets, we derive various results on the probability
distribution of the integration error, which can be compared to
the standard \clt\ for normal stochastic sampling.
In many cases, a Gaussian error distribution is obtained.  

\newpage
\pagestyle{plain}\setcounter{page}{1}

\section{Introduction}
It is widely held that \qmc\ integration, in which the
integration points are distributed more uniformly than in
classical \mc\ integration which uses truly (or approximately)
random points, can lead to potentially much smaller integration
errors for the same amount of effort ({\it i.e.\/} the same number
of integrand evaluations). A number of theorems
are known that relate information on the
fluctuating behaviour of the integrand (such as variation, modulus
of continuity, etc.) and information on the degree of uniformity of
the point set employed (in terms of some quantitative notion
of discrepancy) to the integration error \cite{niederreiter}.
These results, however, do not easily lend themselves to practical
error estimates and moreover, being usually upper limits, may
be too pessimistic in many applications. 

This situation is to be
contrasted to that in classical \mc\ integration: there, one
settles for a {\it probabilistic\/} error estimate, which on the
one hand does not give perfectly certain information but only
confidence levels, but on the other hand can be easily computed
by estimating not only the integral but at the same time the
variance of the integrand. The essential point in this procedure is
the existence of the \clt, which states that for
a large number $N$ of randomly chosen integration points,
the integration error has an approximately Gaussian
distribution with zero mean and a standard deviation related to
the integrand's variance. The estimation of this single parameter
therefore suffices to determine the shape of the error distribution.

In this paper, we attempt to derive results similar to the 
\clt, for the case of \qmc. In previous publications \cite{first,second}
we have argued that such considerations require a definition of
what constitutes an ensemble of $N$-point \qran\ point sets. For truly
random points, this is an easy problem since we may simply assume the
points to be iid uniformly over the integration region.
For \qran\ points the situation is somewhat more subtle. We propose to
use the fact that such more evenly distributed point sets are
generally characterized by a low value of {\it discrepancy\/}:
given {\it some\/} definition of discrepancy (we shall specify one
later on), we restrict ourselves to the set of $N$-point point sets
in which the points are all uniformly iid, but with the additional
condition that the discrepancy has a particular value $s$
(by suitable integration over $s$, we shall of course obtain again the
classical results for truly random points).
We can then study the distribution of the integration error over
this ensemble of point sets.

The lay-out of this paper is as follows. In section 2 we establish
some notation and define our point set ensemble. In section 3 we
derive our main result on the distribution of the integration error,
in terms of a single complex integral. In section 4, we present
explicit results for a particular, simple definition of discrepancy.
In section 5 we attempt to do the same for what we believe
constitutes a realistic discrepancy. In each case we aim at
arriving at an error distribution that depends on only a single
parameter (so that confidence levels for the
integration result can easily be computed), and ultimately, of
course, the ideal Gaussian error distribution.

\section{Notation and definitions}
Our integration region will always be the $D$-dimensional hypercube
$K=[0,1)^D$, containing the point set $X_N=\{x_1,x_2,\ldots,x_N\}$.
Where necessary, we shall denote the individual components of the
vector $x_k$ with Greek indices: so, 
$x_k=x_k\umu=(x_k^1,x_k^2,\ldots,x_k^D)$. Let the integrand be denoted
by $f(x)$; we assume, for simplicity, that the moments
\bq
J_p = \intk dx\;f(x)^p
\eq
exist at least for the first few values of $p$. 
The numerical integral estimate is given by
\bq
S = {1\over N}\suml_{k=1}^N\;f(x_k)\;\;,
\eq
and the integration error $\eta$ is then, of course,
\bq
\eta = S - J_1\;\;.
\eq
It is the probability distribution of $\eta$ over the ensemble of
point sets $X_N$ which is our object of concern here.

We now turn to the definition of a discrepancy. We introduce the
Fourier base of orthonormal function as follows. Starting with $D=1$, 
we define
\bq
u_{2n-1}(x) = \sqrt{2}\sin(2\pi nx)\;\;\;,\;\;\;
u_{2n}(x) = \sqrt{2}\cos(2\pi nx)\;\;,
\eq
for $n=1,2,3,\ldots$, and $u_0(x)=1$. In more dimensions, we 
define vectors $\vn=n\umu=(n^1,n^2,\ldots,n^D)$ with integer, non-negative
components, and write
\bq
\uvn(x) = \prol_{\mu=1}^Du_{n\umu}(x\umu)\;\;.
\eq
We assume that the integrand $f$ can be decomposed into its various
Fourier modes as follows:
\bq
f(x) = \suml_{\vn}v_{\vn}\uvn(x)\;\;,
\eq
from which it immediately follows that
\mpd
\bq
J_1 = v_{(0,0,\ldots,0)}\;\;\;,\;\;\;
V \equiv \suml_{\vn>0}v_{\vn}^2 = J_2-J_1^2 \;\;.
\eq
\mpu
Here and in the following, the notation $\vn>0$ means a sum over all
vectors $\vn$ except the null vector $(0,0,\ldots,0)$. Quadratic
integrability of the integrand requires that the variance $V$, i.e. 
the sum of the $v_{\vn}^2$, converges. \mpl

To each mode with wave vector $\vn$ we associate a {\it strength\/}
$\sivn$. In \cite{first} and \cite{second} we relate these strengths
to a definition of an ensemble of {\it integrands}, by letting
every $v_{\vn}$ be normally distributed with zero mean and width $\sivn$,
but here we do not have to assume a particular such ensemble. 
The definition of (quadratic) discrepancy that we propose to use is
\bq
D_N(X_N) = {1\over N}\suml_{k,l=1}^N\beta(x_k,x_l)\;\;\;,\;\;\;
\beta(x_k,x_l) = \suml_{\vn>0}\sivn^2\uvn(x_k)\uvn(x_l)\;\;.
\eq
An essential property is that
\bq
\intk dx_k\;\beta(x_k,x_l) = \intk dx_l\;\beta(x_k,x_l) = 0\;\;.
\label{betaprop}
\eq
Another important assumption is that of {\it translational invariance}, 
by which the
sines and cosines of each particular wave component have equal strength:
\bqa
\si_{(2n^1,2n^2,\ldots,2n^D)} & = & 
\si_{(2n^1-1,2n^2,\ldots,2n^D)}  =  
\si_{(2n^1,2n^2-1,\ldots,2n^D)} =\nl  
 =\;\;\cdots & = & \si_{(2n^1-1,2n^2-1,\ldots,2n^D-1)} \;\;.
\eqa
One of the consequences of this choice is that $\beta(x_k,x_l)$
only depends on the difference $x_k-x_l$, and therefore
\bq
\beta(0) = \intk dx\;\beta(x,x) = \suml_{\vn>0}\sivn^2\;\;.
\eq
Hence, for truly random points the expected value of the discrepancy is
\bq
\avg{D_N(X_N)} = \suml_{\vn>0}\sivn^2\;\;,
\eq
and of course we assume this sum to converge. 
For the particular point set $X_N$ we are employing, we assume the
discrepancy $D_N(X_N)$ to have a known value $s$, non-negative
by construction. Super-uniform, or
\qran, point sets are distinguished by the fact that $s$ is small
compared to its expectation for random point sets.

We now come to the definition of an ensemble of \qran\ point sets.
We consider it to consist of all point sets $X_N$ that have a value
$s$ of the above discrepancy, but are otherwise unrestricted.
The combined probability density $P_N$ for the $N$ points $x_k$
is then given by
\mpd
\bqa
P_N(s;x_1,x_2,\ldots,x_N) & = & 
{\delta\left(D_N(x_1,x_2,\ldots,x_N)-s\right)\over H_0(s)}\;\;,\nl
H_0(s) & = & \intk dx_1\cdots dx_N\;  
\delta\left(D_N(x_1,x_2,\ldots,x_N)-s\right)\;.
\label{pndef}
\eqa
\mpu
The number $H_0(s)$ serves to normalize the probability density $P_N$: it
is nothing but the probability for a set of truly random points to attain
the value $s$ for its discrepancy. Indeed, we trivially have
\bq
\intl_0^{\infty}ds\;H_0(s)P_N(s;x_1,x_2,\ldots,x_N) = 1\;\;.
\eq

\section{The error distribution}
We now start to work our way towards a \clt\ for \qmc, assuming the
point set $X_N$ to be a member of the ensemble constructed above.
Let $P(s;\eta)$ be the probability density of the integration
error $\eta$ over the ensemble of possible point sets $X_N$. We may
write
\bq
P(s;\eta) = \intk dx_1\cdots dx_N \;P_N(s;x_1,\ldots,x_N)\;
 \delta\left(\eta - S + J_1\right)\;\;.
\label{petadef}
\eq
Using the definition of the Dirac delta distributions
in Eqs.(\ref{pndef},\ref{petadef}) as Fourier integrals, we may write this as
\mpd
\bqa
P(s;\eta) 
& = & {1\over H_0(s)}
\intinfi {dt\over2\pi i}\;{dz\over2\pi i}\;e^{ - z\eta - zJ_1 - ts } M(z,t)\;\;,\nl
M(z,t) & = & \intk dx_1\cdots dx_N\;
\exp\left( {z\over N}\suml_{k=1}^Nf(x_k) 
+ {t\over N}\suml_{k,l=1}^N\beta(x_k,x_l)\right)\nl
 & = & \suml_{m\ge0} {t^m\over m!}\;M_m(z)\;\;,
\eqa
\mpu
where the integration contours for $t$ and $z$ run to the left of
any singularities.

In the spirit of the classical \clt, we must now proceed
to take the asymptotic limit $N\to\infty$ in a careful manner, taking into
account that the dominant part of the $z$ integral comes from
the region where $z$ is of order $\order{\sqrt{N}}$. 
The procedure is most easily illustrated by considering the first
few powers of $t$.
To start, we have
\mpd
\bqa
M_0(z) & = & \intk dx_1\cdots dx_N\;e^{{z}\suml_kf(x_k)/N}
 \;\;=\;\; \ezf ^N
\vphantom{\intk dx_1\cdots dx_N\;e^{{z}\suml_kf(x_k)/N} \ezf ^N}
\nl
& = & \left(1 + {z\over N}J_1 + {z^2\over2N^2}J_2
 + \order{{z^3\over N^3}}\right)^N
\vphantom{\intk dx_1\cdots dx_N\;e^{{z}\suml_kf(x_k)/N} \ezf ^N}
\nl
& = & \exp\left(zJ_1 + {z^2\over2N}(J_2-J_1^2) 
  + \order{{z^3\over N^2}}\right)\;\;.
\vphantom{\intk dx_1\cdots dx_N\;e^{{z}\suml_kf(x_k)/N} \ezf ^N}
\eqa
\mpu
Due to \eqn{betaprop} the next contribution evaluates as follows:
\mpd
\bqa
M_1(z) & = & \intk dx_1\cdots dx_N\;e^{{z}\suml_kf(x_k)/N}
 {1\over N}\suml_{k,l}\beta(x_k,x_l)\nl
& = & \ezf^{N-2}\;{N(N-1)\over2N}\;
 \intk dx_1dx_2\;e^{{z}(f(x_1)+f(x_2))/N}\beta(x_1,x_2)\nl
& & + \ezf^{N-1}\;{N\over N}\;
\intk dx\;e^{{z}f(x)/N}\beta(x,x)\nl
& \sim & M_0(z)\left(\intk dx\;\beta(x,x)
\vphantom{{Z^2\over N}}\;+\right.\nl
& & \hphantom{M_0(z)XX}\left.
 +\; {z^2\over2N}
\intk dx_1dx_2\;f(x_1)\beta(x_1,x_2)f(x_2)\right)\;\;,
\eqa
\mpu
where we have suppressed all subleading terms.
The higher-order terms can easily be worked out: the only
combinations that survive in the limit $N\to\infty$ are
\bqa
C_k & = & \intk dx_1dx_2\cdots dx_k\; 
\beta(x_1,x_2)\beta(x_2,x_3)\cdots\beta(x_{k-1},x_k)\beta(x_k,x_1)\nl
& = & \suml_{\vn>0}\sivn^{2k}\;\;,
\eqa
and
\bqa
F_k & = & \intk dx_1dx_2\cdots dx_kdx_{k+1}\;
f(x_1)\beta(x_1,x_2)\cdots\beta(x_k,x_{k+1})f(x_{k+1})\nl
& = & \suml_{\vn>0}v_{\vn}^2\sivn^{2k}\;\;.
\eqa
These objects come with topological symmetry factors of $2^k/(2k)$ and
$2^k/2$, respectively \cite{first}. 
To leading order in $N$, we can therefore write
\mpd
\bqa
M(z,t) & \sim & M_0(z)\exp\left(
\suml_{k\ge0}C_k{(2t)^k\over2k}
+ {z^2\over N}\suml_{k>0}F_k{(2t)^k\over2}\right)\nl
& = &  M_0(z)\exp\left(
\suml_{k\ge0}\suml_{\vn>0}{(2t\sivn^2)^k\over2k}
+ {z^2\over N}\suml_{k>0}\suml_{\vn>0}{(2t\sivn^2)^kv_{\vn}^2\over2}\right)\nl
& = & M_0(z)\exp\left(
- {1\over2}\suml_{\vn>0}\log(1-2t\sivn^2)
+ {z^2\over2N}\suml_{\vn>0}{2t\sivn^2v_{\vn}^2\over1-2t\sivn^2}
\right)
\eqa
\mpu
Combining everything, we have
\mpd
\bqa
P(s;\eta) & = & {1\over H_0(s)}\intinfi {dz\over2\pi i}\;{dt\over2\pi i}\nl
& & \hphantom{{1\over2\pi i}}\times
\exp\left(-z\eta - ts - {1\over2}\suml_{\vn>0}\log\left(1-2t\sivn^2\right)
 + {z^2\over2N}B(t)\right)\;\;,\nl
B(t) & = & \suml_{\vn>0}{v_{\vn}^2\over1-2t\sivn^2}\;\;.
\eqa
\mpu
The $z$ integral converges provided $\mbox{Re}B(t)>0$, which 
certainly holds if $1-2\sivn^2\mbox{Re}t >0$ for all $\vn$.
Performing the $z$ integration, we arrive at our master formula:
\mpd
\bqa
P(s;\eta) & = & {1\over H_0(s)}
\intinfi {dt\over2\pi i}\;\sqrt{N\over2\pi B(t)}\nl
& & \hphantom{\sqrt{N\over2\pi} {1\over H}}\times
\exp\left(-ts-{1\over2}\suml_{\vn>0}\log\left(1-2t\sivn^2\right)
-{\eta^2N\over2B(t)}\right)\;\;.
\label{master}
\eqa 
\mpu

We see that, for the types of discrepancy discussed here, the
error distribution is symmetric around $\eta=0$. Its precise form,
however, will depend on our choice for the $\sivn$. As we have said,
a particular such choice reflects our belief about which kind of
function class our actual integrand is a typical member of: but it
must be realized that we are, in fact, allowed to take {\it any\/}
choice for the $\sivn$ that satisfies $\sum\sivn^2<\infty$. A choice
that does not `fit' the behaviour of $f(x)$ too well will just
result in a somewhat worse error estimate: but the error distribution
itself is only based on our assumption on the ensemble of point sets $X_N$,
and not on any assumption about the integrand apart from its
quadratic integrability.

From \eqn{master} a number of results immediately follow. In the
first place, we can recover the case of truly random point sets
by simply averaging
over all possible values of $s$, with the appropriate probability
distribution\mpl $H_0(s)$: this immediately leads to
\mpd
\bq
\intl^\infty_0 ds\;H_0(s)\;P(s;\eta) = \sqrt{{N\over2\pi V}}
\exp\left(-{\eta^2N\over2V}\right)\;\;\;, 
\eq
\mpu
which is the standard \clt.
Another result comes from the normalization of $P(s;\eta)$: upon integrating
over $\eta$ we find
\bq
H_0(s) = \intinfi {dt\over2\pi i}\;
\exp\left( - ts - {1\over2}\suml_{\vn>0}\log\left(1-2t\sivn^2\right) \right)\;\;,
\label{h0def}
\eq
in accordance with Ref.\cite{first}. A final observation to be made is
that the error $\eta$ only occurs in the combination $\eta^2N$. From this
it immediately follows that, all other things being equal, the error
will only decrease as $1/\sqrt{N}$. Any improved rate of convergence
is therefore {\it solely\/} due to a decrease of the
discrepancy value $s$ with $N$.

\section{A simple model: uniform strengths}
The first, and simplest, model that we shall consider is that
where $2M$ of the $\sivn^2$ are equal to $1/2M$, and all the other
ones vanish. It is natural to take for the nonzero modes the ones with
the lowest frequencies ({\it i.e.\/} small values of the components
of $\vn$), but this is not necessary. As mentioned above, the
choice of $\sivn$ only establishes which modes are {\it covered},
that is, enter in
the computation of the discrepancy: a general integrand will, of course
have modes with different frequencies, which are not covered. We
therefore write
\bq
V = \suml_{\vn>0}v\svn^2 = V_1 + V_2\;\;,
\eq
where $V_1$ contains the $2M$ covered modes, 
for which $\sivn\ne0$, and $V_2$
contains all the other, uncovered, ones. 
The larger $V_1$ is with respect to $V_2$,
the better our discrepancy model `fits' the integrand. 
We immediately have
\mpd
\bqa
{1\over2}\suml_{\vn>0}\log\left(1-2t\sivn^2\right) & = & 
M\log\left(1-{t\over M}\right)\;\;,
\vphantom{\sqrt{M\over2\pi}\exp\left(-{M(s-1)^2\over2}\right)}
\nl
B(t) & = & V_1/\left(1-{t\over M}\right) + V_2\;\;,
\vphantom{\sqrt{M\over2\pi}\exp\left(-{M(s-1)^2\over2}\right)}
\nl
H_0(s) & = & {M^M\over\Gamma(M)}s^{M-1}e^{-Ms}
\vphantom{\sqrt{M\over2\pi}\exp\left(-{M(s-1)^2\over2}\right)}
\nl
& \sim & 
\sqrt{M\over2\pi}\exp\left(-{M(s-1)^2\over2}\right)
\;\;,
\eqa
\mpu
where the last line holds for large $M$.
Both the form of $H_0(s)$ and that of $\beta(x_k,x_l)$ for
this model are given in \cite{second}; by construction,
the expected discrepancy for truly random points is 
$\avg{s}=1$. The master formula now becomes
\mpd
\bqa
P(s;\eta) & = & {\Gamma(M)\over M^{M-1}}
\intinfi {dx\over2\pi i}\;\sqrt{N\over2\pi(V_2 + sV_1/x)}\nl
& & \hphantom{{\Gamma(M)\over M^{M-1}}}
\times \exp\left( Mx - M\log x - {\eta^2N\over2(V_2+sV_1/x)}\right)\;\;,
\eqa
\mpu
where we have written $x\equiv s(1-t/M)$. \mpl Consequently the integration
contour must cross the positive real axis. Two special cases can
immediately be derived from this. In the first place, suppose that
we had chosen the nonzero $\sivn$ in a very bad way, such that
$V_1=0$: that is, the integrand consists only of uncovered modes.
It then follows immediately that
\mpd
\bq
\left.P(s;\eta)\right|_{V_1=0} = 
\sqrt{N\over2\pi V_2}\exp\left(-{\eta^2N\over2V_2}\right)\;\;,
\eq
\mpu
which is the standard \clt. In this case, nothing is really lost,
and the error estimate is just as good (or bad) as in classical \mc.
On the other hand, if the integrand consists only of covered modes, so that
$V_2=0$, we find after some straightforward manipulations:
\mpd
\bqa
P(s;\eta) & = & \xi(M)\;\sqrt{N\over2\pi V_1s}\;
\left(1-{\eta^2N\over2V_1sM}\right)^{M-3/2}\;\;,\nl
\xi(M) & = & {4^{M-1}\over\sqrt{M\pi}}{\Gamma(M)^2\over\Gamma(2M-1)}
\;\;{=}\;\;1+\order{{1\over M}}\;\;,
\label{v2zerocase}
\eqa
\mpu
with the strict constraint $\eta^2N<2V_1sM$. This follows from the fact
that, if this inequality is violated, the complex integration contour
for $x$ can be closed to the right, where the integrand has no
singularities; for the same reason \cite{second}, $H_0(s)$
vanishes for $s<0$. Note that, for this particular discrepancy,
$s$ can actually vanish: this happens in one dimension,
if the point set is equidistant and $N>M$. In that case, $\eta$ is
always zero, so that the function is integrated exactly. This is
just another instance of the Nyqvist theorem \cite{nyq}.

For general $V_1$ and $V_2$, we may consider the case where $M$
becomes large. The integral can then be approximated by the
saddle-point method. The saddle point is located at
$x=1+\order{1/M}$, and we find
\bq
P(s;\eta) \sim \sqrt{{N\over2\pi(V_1s+V_2)}}
\exp\left(-{\eta^2N\over2(V_1s+V_2)}\right)\;\;.
\label{simplemodelclt}
\eq
Again, we recover a Gaussian limiting distribution; its width
is no longer parameterized by $V=V_1+V_2$ but rather by
$V_1s+V_2$: the information we have gathered by computing the
discrepancy $s$ is seen to result in a reduced error, depending on
how much of the fluctuating behaviour of the integrand is actually
covered by the modes entering in the discrepancy. The limit of
large $M$ is actually justified by a self-consistency argument: the
error distribution (\ref{simplemodelclt}) heavily suppresses the
region $\eta^2N\gg2(V_1s+V_2)$, so that (as can also be gleaned from
\eqn{v2zerocase}) $M$ does not actually have to be a huge number
for the saddle-point approximation to work. Note, moreover, that
if we only allow the lowest frequency mode in each dimension, that is,
only $n\umu=0,1,2$ for each component of $\vn$, $M$ already
equals $(3^D-1)/2$ which grows very rapidly with increasing $D$.
The upshot of this (admittedly simple-minded, but nevertheless
possible) model is: first, that we may hope for an error distribution 
which tends to a Gaussian (especially in high dimension), and, secondly,
the width of this distribution depends on the discrepancy $s$ in
a manner which depends on the degree in which the relevant modes of the
integrand correspond to those used in the evaluation of the discrepancy.
We conjecture that these two conclusions will persist in more
realistic models of discrepancy. 

\section{A more realistic model: one dimension}
The model of discrepancy discussed above has the advantages both
of simplicity and dimensionality-independence: but it may not be
altogether too realistic, in particular because covered modes with high
frequency are assumed to have the same strength as those with low
frequency. An alternative, which we discuss now, covers all modes, but
with strengths that decrease with increasing frequency. For
simplicity, we start with $D=1$. We shall take
\mpd
\bq
\si_{2n} = \si_{2n-1} = {1\over n}\;\;\;,\;\;\;n=1,2,3,\ldots,
\eq
\mpu
just the same as in \cite{second}. For truly random points
we have, then, $\avg{s}=\pi^2/3$, and we shall assume that we
have at our disposal a point set with a discrepancy value $s$ 
much lower than this average.
First of all, we compute $H_0(s)$ for this small $s$. In Ref.\cite{second},
we performed an exact calculation, but here we shall settle for
a more simple-minded saddle-point approximation. We assume that
the $t$ integral in \eqn{h0def} is saturated by a saddle-point
lying at $t=-a^2/2$, that is,
\mpd
\bqa
H_0(s) & = & \intinfi {dt\over2\pi i}\;e^{\phi(t)}\nl
& \sim & {\exp\left({\phi(-a^2/2)}\right)\over\sqrt{2\pi\phi''(-a^2/2)}}\;\;,
  \vphantom{{1\over2\pi i}\intinfi dt\;e^{\phi(t)}}
\nl
\phi(t) & = & -st - \suml_{n>0}\log\left(1-{2t\over n^2}\right)\;\;,
  \vphantom{{1\over2\pi i}\intinfi dt\;e^{\phi(t)}}
\nl
\phi(-a^2/2) & = & {sa^2\over2} - \pi a + \log(2\pi a) 
                + \order{{1\over a}}\;\;,
  \vphantom{{1\over2\pi i}\intinfi dt\;e^{\phi(t)}}
\nl[-3mm]
\phi'(-a^2/2) & = & -s + {\pi\over a} + \order{{1\over a^2}}\;\;\equiv0\;\;,
  \vphantom{{1\over2\pi i}\intinfi dt\;e^{\phi(t)}}
\nl[-3mm]
\phi''(-a^2/2) & = & {\pi\over a^3} + \order{{1\over a^4}}\;\;.
  \vphantom{{1\over2\pi i}\intinfi dt\;e^{\phi(t)}}
\eqa
\mpu
The saddle point is seen to correspond to $a\sim\pi/s$ which
is large for small values of $s$, thus justifying the neglect of
higher orders in $1/a$. The resulting form for $H_0$ is
($s\ll{\pi^2/3}$)
\mpd
\bq
H_0(s) \,\sim\, 
{\pi^2\sqrt{2\pi}\over s^{5/2}}\exp\left(-{\pi^2\over2s}\right)\;\;, 
\eq
\mpu
in agreement with the corresponding limit of
the exact result from \cite{second}.

For the evaluation of the error distribution $P(s;\eta)$ we
must now also compute $B(t)$, which involves the unknown
coefficients $v_n$ of the integrand. It is certainly too crude,
but nonetheless instructive, to study the simple case where
$$
v_n^2 = \si_n^2\;\;\;,\;\;\;n=1,2,3,\ldots.
$$
In that case, we have
\bq
B(t) = 2\suml_{n>0}{1\over n^2-2t}
= 2\suml_{n>0}{1\over n^2+a^2}
= {\pi\over a} + \order{{1\over a^2}}\;\;,
\label{btavg}
\eq
where $a$ has now to be determined anew for the saddle point in the $t$
integration of \eqn{master}. It is seen to be equal to
\bq
a \sim {\pi\gamma\over s}\;\;\;,\;\;\;
\gamma = 1 + {\eta^2N\over2\pi^2}\;\;.
\eq
Performing the saddle integral we arrive at
\bqa
P(s;\eta) & \sim & \sqrt{N\over2\pi s}\;\gamma^{5/2}\;
\exp\left(-{\pi^2\over2s}(\gamma^2-1)\right)\nl
& \sim & \sqrt{N\over2\pi s}\exp\left(-{\eta^2N\over2s}\right)\;\;.
\eqa
This last, Gaussian, central limit is self-consistently justified
from the fact that it implies $\eta^2N=\order{s}$ which is indeed
small by assumption. 
\mpd
Note that we may write for this case 
(see \eqn{simplemodelclt}):
$$
s = V{s\over\left<s\right>}\;\;.
$$
\mpu

What, now, happens for more general $v_n$? One answer is to
assume that, since the integrand must be quadratically integrable,
the sum $\sum v_n^2$ must converge; if we also assume that it
has no exceptionally strong higher modes, it is reasonable to
write
$$
v_{2n-1}^2 + v_{2n}^2 = {C\omega_n\over n^2}\;\;,
$$
where $C$ is a constant, and the $\omega_n$ are numbers that
are not too different from unity. Not rigorously, but at least
reasonably, we may then write
\bq
B(t) = \suml_{n>0}{C\omega_n\over n^2+a^2}
\sim {C\pi\over a} + \order{{1\over a^2}}\;\;,
\eq
leading to
\bq
P(s;\eta) \sim \sqrt{N\over2\pi sC}\;
\exp\left(-{\eta^2N\over2sC}\right)\;\;.
\eq
The essential point here is that the deviations of the individual
$\omega_n$ from unity can give rise, in $B(t)$ to contributions
that are of order $\order{1/a^2}$, and not of order $\order{1/a}$.
Another argument leading to the same conclusion is to compute the
moments of $B(t)$ over the ensemble of integrands described in
Refs.\cite{first,second}: the $v_n$ are assumed to be normally
distributed around zero with standard deviation $\si_n$. The
expectation of $B(-a^2/2)$ is then, of course, just the
result of \eqn{btavg}, but its {\it variance\/} goes as $\order{1/a^3}$.
If $a$ increases (for decreasing $s$), the probable values for $B(t)$
therefore cluster together more and more
closely around the expectation value, again justifying our
approximations.

A last example in this context is that of an integrand that has
only a single mode, with frequency $k$, so that only
$v_{2k}$ and $v_{2k-1}$ are non-vanishing, and we have
\mpd
\bq
B(t) = {Vk^2\over k^2-2t}\;\;.
\eq
\mpu
We immediately find that 
\bq
\hat{s} \equiv s - {\eta^2N\over Vk^2} > 0\;\;,
\eq
by the same arguments as above\mpl; and, for $\eta$ values smaller than
this limit, we may again apply the saddle-point method to find
\bqa
P(s;\eta) & \sim & \sqrt{{N\over2\pi V}}\;
\left({s\over\hat{s}}\right)^{5/2}\sqrt{1+{\pi^2\over\hat{s}^2k^2}}
\exp\left(-{\eta^2N\over2V}+{\pi^2\over2s}-{\pi^2\over2\hat{s}}\right)\nl
& \sim & \sqrt{{N\over2\pi V}\left(1+{\pi^2\over k^2s^2}\right)}
\exp\left(-{\eta^2N\over2V}\left(1+{\pi^2\over k^2s^2}\right)\right)\;\;,
\eqa
where the last line holds if $\hat{s}$ and $s$ are close in value.
In this limit, again a Gaussian distribution is obtained, with 
variance $V/(1+\pi^2/k^2s^2)$. Note that the error improvement now
not only depends on the smallness of $s$ but also on the number $k$;
this is reasonable because the mode with frequency $k$ enters in this
particular discrepancy with a factor $1/k^2$ so that, when $k$
is large, a small value of $s$ does not tell us too much about how well
the $k$th mode is integrated by the point set.

\mpd
\section{Conclusions and outlook}
We have shown that we can define a Central Limit for the case of 
Quasi-Monte Carlo using a suitable definition of the discrepancy.
A master-formula was derived for the error-distribution density
over point sets with a fixed discrepancy.

We have given two simple examples of problem classes and their  
error-distribution densities. These results indicate that 
the expected error will improve if low-discrepancy
point sets are used to evaluate integrals.

We would like to extend these results to more realistic and more
dimensional cases. An explicit result seems to be too far-fectched,
at the moment, but it might be possible to use saddle-point methods to
derive similar results for more realistic cases.
\mpu


\end{document}